\documentclass[aps,prl,twocolumn,showpacs]{revtex4}
\usepackage{epsfig}
\usepackage{graphicx}
\usepackage{bm}
\usepackage{amssymb}
\usepackage{amsmath}
\usepackage{amsthm}
\begin{document}
\title{Dynamical correlations near dislocation jamming}
\author{Lasse Laurson$^{1}$, M.-Carmen Miguel$^2$, and Mikko J. Alava$^3$}
\affiliation{$^1$ ISI Foundation, Viale S. Severo 65, 10133 Torino, Italy}
\affiliation{$^2$ Departament de F\'\i sica Fonamental, Facultat de F\'isica,
Universitat de Barcelona, Diagonal 647, 08028 Barcelona, Spain}
\affiliation{$^3$ Aalto University, Department of Applied Physics,
PO Box 14100, 00076 Aalto, Espoo, Finland}
\begin{abstract}
Dislocation assemblies exhibit a jamming or yielding transition at a
critical external shear stress value $\sigma=\sigma_c$. Nevertheless 
the nature of this transition has not been ascertained. Here we study the
heterogeneous and collective nature of dislocation dynamics within a 
crystal plasticity model close to $\sigma_c$, by considering the 
first-passage properties of the dislocation dynamics. 
As the transition is approached in the moving phase, 
the first passage time distribution exhibits scaling, and a related
peak {\it dynamical} susceptibility $\chi_4^*$ diverges as $\chi_4^* \sim
(\sigma-\sigma_c)^{-\alpha}$, with $\alpha \approx 1.1$. We relate this
scaling to an avalanche description of the dynamics. While the
static structural correlations are found to be independent of the
external stress, we identify a diverging dynamical correlation
length $\xi_y$ in the direction perpendicular to the dislocation
glide motion.
\end{abstract}
\maketitle

The mechanical behavior of crystalline solids subject to an external
shear stress $\sigma$ is controlled by the existence of a finite
yield stress $\sigma_c$: for stresses below $\sigma_c$ only elastic
or reversible deformation of the material takes place, while
sustained plastic or irreversible deformation mediated by
dislocation motion is observed for $\sigma>\sigma_c$. The basic
phenomenology of two states with different rheology is the same also
for many amorphous materials ranging from foams to amorphous and
granular media, even though the microscopic mechanisms responsible
for a transition to "flow" are not yet well understood. The concept
of {\em jamming} has been introduced to comprehend the observed
phenomenology: In a finite region of the parameter space (spanned by
control variables as temperature, density and an external force),
due to a "cage effect", or self-induced constraints on the motion of
the system constituent particles, the phase space dynamics gets
restricted, i.e. jammed \cite{LIU-01,BIR-07,OHE-02,MIG-06}. In
particular for two-dimensional ensembles of crystal dislocations a
yielding transition at a finite $\sigma_c$ has been established, 
even in the absence of any external sources of disorder \cite{MIG-02}. 
The crucial mechanism here is the generation of effective disorder 
in the local stress field, due to constrained dislocation motion 
by the combined effect of the slip geometry and the 
long range anisotropic dislocation-dislocation interactions, which 
are also manifest in the formation of typical, metastable elementary 
structures such as dislocation dipoles and walls.

On approaching the jamming transition from the moving phase, the dynamics 
generally becomes increasingly heterogeneous, and a growing dynamical
correlation length characterizing correlation across intervals of
both space and time has been observed for granular systems
\cite{KEY-07,ABA-07}, as well as on approaching the glass transition
of molecular liquids and colloidal suspensions \cite{BER-05}. In
this work, we consider dislocation jamming in a similar vein. Known
features of the 2d discrete dislocation dynamics (DDD) model studied 
here are a relaxation of the strain rate or an Andrade law close 
$\sigma=\sigma_c$, following a power law $\dot{\epsilon} \sim t^{-\theta}$, 
$\theta \approx 2/3$ \cite{MIG-02}. The steady state strain rate 
displays a non-linear rheological behavior, $\dot{\epsilon} \sim
(\sigma-\sigma_c)^{\beta}$, with $\beta \approx 1.8$ \cite{MIG-02}. 
The model also reproduces the experimentally observed scale free size 
distributions of avalanches of plastic deformation \cite{MIG-01}. 
It is also similar to the DDD models used in the context of many 
engineering applications \cite{NIC-03}.
The question is now, 
whether one can establish the jamming/yielding transition as a second 
order phase transition. This would imply the existence of a length scale,
the correlation length, diverging at the transition point. The
dislocation system is particular, since here the transition is of a
non-equilibrium kind: the temperature is included only indirectly
via the dislocation mobility and is thus irrelevant.

The avalanches exhibited by dislocation assemblies \cite{MIG-01} are 
suitable candidates for fundamental ``dynamical heterogeneities'' or 
localized events.
We define appropriate statistical quantities to characterize the dynamics
via time-dependent {\em first passage probabilities} which measure
the likelihood that a dislocation becomes
liberated from the confining stress field over an observation scale.
The first passage time distributions are found to display scaling close to the
transition, and
analogously to other systems exhibiting
jamming, the peak value of a {\it dynamical susceptibility} and a
{\it dynamical correlation length} grow on approach to jamming and
diverge at $\sigma=\sigma_c$. At the same time {\it static} 
correlations describing the dislocation structures are found to be 
virtually independent of the external stress. We also connect the 
results to an avalanche description of the dynamics.

\begin{figure}[t]
\includegraphics[angle=-90,width=7.5cm,clip]
{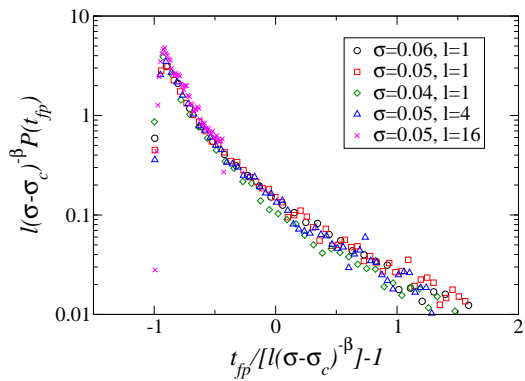}
\caption{(color online) Scaled first passage time distributions according to Eq. (\ref{eq:fp}),
for various values of $\sigma$ and $l$.
The value of the exponent $\beta$ is $\beta=1.8$.}
\label{fig:fpt}
\end{figure}

The DDD model can be considered to represent a $2d$ cross section ($xy$
plane) of a $3d$ single crystal with a single slip geometry
\cite{MIG-02,MIG-01,MIG-01_2}. The
dislocations are taken to be straight parallel edge dislocations, with
the dislocation lines along the $z$ axis. The $N$ point-like dislocations
glide along directions parallel to their Burgers vectors $\vec{b}_n=s_n b
\vec{u}_x$, with $s_n = \pm 1$ and $\vec{u}_x$ the unit vector in the
$x$ direction. Equal numbers of dislocations with positive and negative
Burgers vectors are assumed. Dislocation climb is not considered for
simplicity. The dislocations interact with each other through their
anisotropic long-range stress fields,
\begin{equation}
\sigma_s(\vec{r}) = Db \frac{x(x^2 - y^2)}{(x^2+y^2)^2},
\end{equation}
where $D=\mu/2\pi(1-\nu)$, with $\mu$ the shear modulus and $\nu$ the
Poisson ratio of the material. The dynamics is taken to be overdamped
with a linear force-velocity relation, resulting in the equations of
motion
\begin{equation}
\frac{\chi_d^{-1}v_n}{b} = s_n b \left[\sum_{m \neq n}
s_m\sigma_s(\vec{r}_{nm}) + \sigma \right],
\end{equation}
where $v_n$ is the velocity of the $n$th dislocation, $\chi_d$ is the
dislocation mobility, $s_n$ is the sign of the Burgers vector of the
$n$th dislocation, and $\sigma$ is the applied external shear stress.
The long range interaction forces are
computed by imposing periodic boundary conditions in {\it both} $x$ and $y$
directions. The equations of motion are then integrated numerically with
an adaptive step size fifth order Runge-Kutta algorithm. This is done
in dimensionless units by measuring lengths in units of $b$, times in
units of $1/(\chi_d Db)$, and stresses in units of $D$. If
the distance between two dislocations of opposite sign gets smaller than
$2b$, they are removed from the system, a process reminiscent of the
annihilation reaction occurring in real plastically deforming crystals.
The simulations are started from a random initial configuration of $N_0$
dislocations within a square cell of linear size $L$. The system is
first let to relax with $\sigma=0$,
in the absence of external stresses, 
until it reaches
a metastable dislocation arrangement. During this initial relaxation,
a significant fraction of the dislocations get annihilated. We consider
here systems of linear size $L=200b$ and $L=300b$, with
$N_0=1600$ and $N_0=3600$, respectively. After the relaxation,
the external stress is switched on, and the evolution of the system
is monitored. For $\sigma>\sigma_c \approx 0.025$ \cite{sigmac}, the system eventually
reaches a steady state, with the strain rate $\dot{\epsilon} \sim \sum_n b_n v_n$
fluctuating around a constant mean value. This steady state consists of $N=350-400$ 
dislocations for $L=200b$, and $N=800-900$ for $L=300b$.

\begin{figure}[t]
\includegraphics[width=4.8cm,angle=-90,clip]
{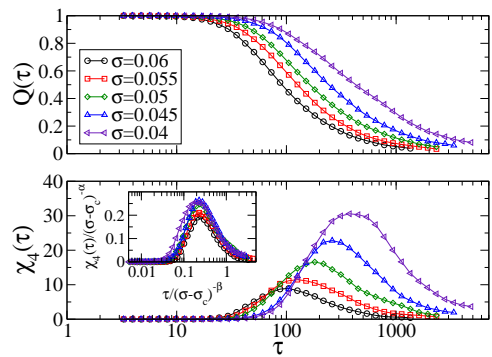}
\caption{(color online) {\bf Top:} The average self-overlap order parameter $Q(l,\tau)$
for various values of $\sigma$ as a function $\tau$, for $l=1$. {\bf Bottom:}
The corresponding four-point dynamic susceptibilities $\chi_4(l,\tau)$.
The inset shows evidence of a data collapse of $\chi_4(\tau)$, obtained by
rescaling the data with the scaling forms of $\langle t_{fp} \rangle$ and $\chi_4*$,
with $\alpha=1.1$ and $\beta=1.8$.}
\label{fig:qchi}
\end{figure}

To characterize the heterogeneous dynamics we first consider the
first passage time distribution of dislocations in the moving steady
state. The first passage time $t_{fp}$ is defined as the time at
which a dislocation first moves across a distance $l$ from some
initial position, and thus becomes liberated from a confining stress 
field at this scale. Given the observed non-linear stress dependence of
the steady state strain rate \cite{MIG-02}, the mean first passage
time is expected to behave like $\langle t_{fp} \rangle \sim
l/\langle v \rangle \sim l(\sigma-\sigma_c)^{-\beta}$. Assuming
$\delta t_{fp} \sim \langle t_{fp} \rangle$, one can write
\begin{equation}
P(t_{fp};l,\sigma) = \frac{1}{l(\sigma-\sigma_c)^{-\beta}} \mathcal{F}
\left[ \frac{t_{fp}-l(\sigma-\sigma_c)^{-\beta}}{l(\sigma-\sigma_c)^{-\beta}} \right].
\label{eq:fp}
\end{equation}
Fig. \ref{fig:fpt} shows a data collapse according to Eq.
(\ref{eq:fp}) with $\beta = 1.8$, in good agreement 
with the scaling of strain rate with the applied
stress \cite{MIG-02}. The scaling function $\mathcal{F}(x)$ exhibits
a maximum and a tail which does not contribute much to the typical
escape characteristics. 

A cumulative version of $P(t_{fp})$ has been called
the instantaneous self-overlap
order parameter \cite{ABA-07},
\begin{equation}
Q_t(l,\tau) = \frac{1}{N} \sum_{n=1}^N w_n,
\end{equation}
where $N$ is the number of dislocations and $w_n$ equals $1$ if the
displacement of dislocation $n$ remains less than $l$ across the time
interval $t \rightarrow t+\tau$ and equals 0 otherwise. The first and
second moments
\begin{eqnarray}
Q(l,\tau) & = & \langle Q_t(l,\tau) \rangle \\
\chi_4(l,t) & = & N[\langle Q_t(l,\tau)^2\rangle-\langle Q_t(l,\tau)\rangle^2]
\label{eq:chi4def}
\end{eqnarray}
of $Q_t(l,\tau)$ (calculated from sample to sample fluctuations
of $Q_t(l,\tau)$) are then used to characterize the dynamics.
As is customary for similar problems, we consider here only samples
that are still active at the observation time \cite{DIC-05}.
By construction, $Q(l,\tau)$ decays from one to zero as a function of
$\tau$, while the {\it four-point dynamic susceptibility} 
$\chi_4(l,\tau)$ exhibits a peak at an intermediate $\tau=\tau_4^*$, 
and vanishes for both early and late times, see Fig. \ref{fig:qchi}. 
$Q^*=Q(l,\tau_4^*) \approx 0.5$ is nearly a constant. The peak value 
$\chi_4^*$ of $\chi_4(l,\tau)$ depends on the proximity of the critical 
stress $\sigma_c$, and has been argued to be related to the typical 
number of particles (here dislocations) in a correlated fast-moving 
domain or heterogeneity \cite{ABA-07}.

\begin{figure}[t]
\includegraphics[width=5.0cm,angle=-90,clip]
{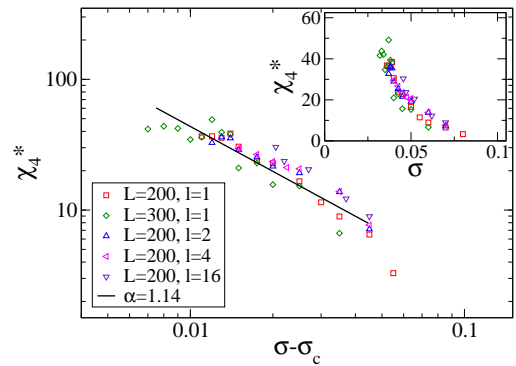}
\caption{(color online) The peak value $\chi_4^*$ of the dynamic susceptibility as a function of
the distance from the critical stress for different system sizes $L$ and length scale 
parameters $l$. The solid line corresponds to a power law $\chi_4^* \sim 
(\sigma-\sigma_c)^{-\alpha}$, with $\alpha=1.14$.
The inset shows the same data as a function of $\sigma$ with linear axis scale.}
\label{fig:chi4div}
\end{figure}

In the simulations, we observe a growing $\chi_4^*$ as $\sigma_c$ is
approached from above, with an apparent divergence at $\sigma =
\sigma_c \approx 0.025$, see Fig. \ref{fig:chi4div}. 
For external stresses $\sigma \leq 0.07$, the stress dependence of $\chi_4^*$ can
be characterized by a power law $\chi_4^* \sim (\sigma-\sigma_c)^{-\alpha}$, 
with $\alpha \approx 1.1$. This result does not exhibit any
significant $l$-dependence. The scaling of $\chi_4^*$ can be connected 
to the avalanche dynamics close to the transition by directly 
computing the variance of $Q(l,\tau)$ at $\tau_4^*$. Consider
the distribution of avalanche sizes $s$ with a $\sigma$-dependent cut-off,
$P(s)=s^{-\tau}f[s/(\sigma-\sigma_c)^{-\gamma}]$. By assuming that the
number of dislocations $n$ swept to a distance $l$ during an avalanche obeys
$\langle n \rangle \sim \langle s \rangle$, and using the condition
$M\langle n \rangle = N/2$ (arising from the observation that $Q^*\approx 0.5$)
to estimate the number $M$ of independent avalanches 
occurring up to the delay time $\tau_4^*$, we obtain 
$\chi_4^*=N(\delta Q^*)^2 \sim N(\sqrt{M}\langle n \rangle/N)^2 \sim 
\langle n \rangle/2 \sim (\sigma-\sigma_c)^{-\gamma(2-\tau)}$.
As $\langle n \rangle$ grows on approaching $\sigma_c$, the number
$M$ of heterogeneities contributing to $\chi_4^*$ decreases:
an estimate from the numerics for $L=300b$ leads to $M \approx 19$ 
for $\sigma=0.06$ and $M \approx 6$ for $\sigma=0.037$.
Given that the avalanche size
exponent is known to have a value close to the mean field 
$\tau \approx1.5$ \cite{MIG-01_2}, $\alpha \approx 1$ would indicate 
$\gamma \approx 2$, again consistent with the mean field -like scenario 
\cite{ZAI-05}. 

\begin{figure}[t]
\includegraphics[width=7.0cm,clip]
{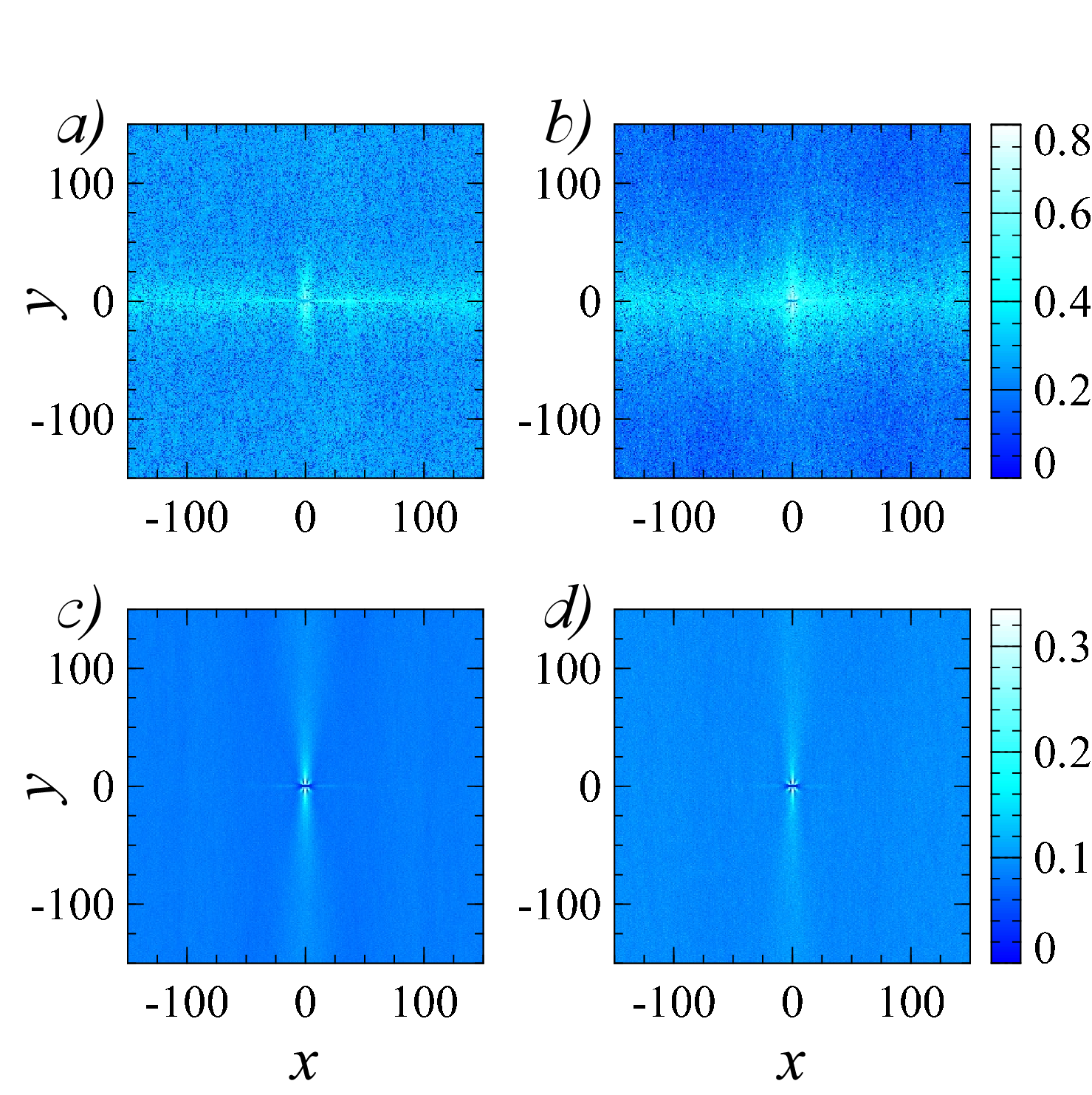}
\caption{(color online) The dynamic and static correlation functions
from a system of size $L=300b$ and $l=1$, with the origin in the middle 
of the figures. In {\bf a)} and {\bf b)} we show the dynamic 
correlation functions $g_d(\vec{r},l=1)$ for $\sigma=0.06$ and
$\sigma=0.04$, respectively. In {\bf c)} and {\bf d)}, the
static correlation function $g(\vec{r})$ for the same stress 
values ($\sigma=0.06$ in {\bf c)} and $\sigma=0.04$ in {\bf d)})
are displayed.}
\label{fig:maps}
\end{figure}

\begin{figure}[t!]
\includegraphics[width=5.0cm,angle=-90,clip]
{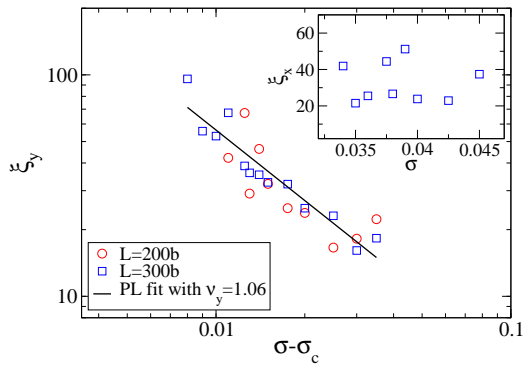}
\caption{(color online) {\bf Main figure:}
The dynamic correlation length $\xi_y$ in the $y$-direction, exhibiting a power law
divergence close to $\sigma_c$. {\bf Inset:} The dynamic correlation length $\xi_x$ in the
$x$ direction. The length scale parameter $l=1$ is used.}
\label{fig:xi}
\end{figure}

To further explore the possibility of identifying a divergent 
non-equilibrium correlation length on
approach to dislocation jamming, we consider the conditional probability
$g_d(\vec{r},l)=P[w(l,\tau_4^*,\vec{r})=0|w(l,\tau_4^*,0)=0]$ that a dislocation 
at a position $\vec{r}$ has
moved at least a distance $l$ up to a delay time $\tau_4^*$, given that a dislocation at the
origin has done so. $g_d(\vec{r},l)$ provides a measure of the spatial structure of the typical 
heterogeneities contributing to $\chi_4^*$. Figs. {\ref{fig:maps}} a) and b) show examples of this
dynamical correlation function for stress values $\sigma=0.06$ and $\sigma=0.04$,
respectively. We focus here for simplicity on the case $l=1$. Notice that closer to jamming,
correlations clearly extend further in
the $y$ direction, while we could not identify a clear
trend in the $x$ direction. At the same
time, the static density-density correlation function $g(\vec{r})$, computed from snapshots
of the dislocation configurations in the steady state, does not show any significant
dependence on $\sigma$ \cite{ZAI-01,correlsign}. To estimate the dynamical correlation length, we fitted an
exponential function $\exp (-y/\xi_y)$ to the $g_d(\vec{r})$-data averaged over
a narrow strip along the $y$-direction. This leads to a divergent stress dependence of
$\xi_y \sim (\sigma-\sigma_c)^{-\nu_y}$, with $\nu_y \approx 1$, see
Fig. \ref{fig:xi}. A similar analysis in the $x$-direction is less conclusive: due to the
tendency of dislocations to form wall-like structures, the statistics is significantly
worse in the $x$ direction, and no clear trend is visible for the evolution of $\xi_x$
with $\sigma$ (inset of Fig. \ref{fig:xi}): $\xi_x$ appears to have a relatively large
stress-independent value around $\xi_x \approx 35b$. 
Recall that in $3d$ DDD simulations \cite{CSI-07},
avalanches have been observed to be characterized by a lamellar shape,
something that could be related to this large and stress-independent value of $\xi_x$.
Notice also that the observed scaling of $\xi_y$ is consistent with the mean field cut-off 
scaling of the avalanche size distribution, assuming that $\xi_y$ measures the linear size of
a {\it typical} avalanche in the $y$ direction.

In summary, we have shown that the dislocation jamming transition is
accompanied by a divergent dynamic susceptibility and a divergent correlation
length. 
This scaling, which we relate to the avalanche dynamics of the system,
is purely dynamical in nature: the static correlation
function characterizing the dislocation structures does not exhibit any
significant dependence on $\sigma$.
The observed exponent values $\alpha \approx \nu_y \approx 1$
indicate that the heterogeneities are compact and characterized by
a single diverging correlation length in the $y$ direction,
thus clearly demonstrating that the dislocation jamming transition is 
indeed a true non-equilibrium second order phase transition.
It would be interesting to test similar avalanche-based ideas 
also for plasticity of amorphous solids \cite{LEM-09}.
Notice that despite similarities with other relevant systems such as granular 
media, there are also significant differences:
In the dislocation system the dynamics stops due to the
formation of various metastable dislocation structures (dipoles, walls, and
more complicated multipolar structures) arising from the
combination of anisotropic long-range dislocation-dislocation interactions
and restricted dislocation motion along a single slip direction. On the
other hand, e.g. in granular systems jamming takes place when grains get
stuck by short range contact forces within {\it cages} formed by their nearest
neighbors, a process that is not accompanied by any large scale grain
structures. The observation that in the present case the results seem to be
independent of $l$ close enough to the
transition suggest that contrary to many other systems 
where $l$ needs to be tuned to a specific value related to an 
underlying ``cage size'', such a typical microscopic length does not exist 
here. This indicates that the same scale free property would apply also for the 
force landscape experienced by the dislocations, with the first passage 
properties related to the time it takes for the force landscape to deform on 
the observation scale, due to the avalanche dynamics.

{\bf Acknowledgments}. D.J. Durian and M. Zaiser are thanked for useful comments.
We acknowledge the support of the Academy of Finland (via 
a post-doctoral grant and via the Center of Excellence -program) and 
the European Commissions NEST Pathfinder programme TRIGS under 
contract NEST-2005-PATH-COM-043386.
MCM acknowledges the support of the Ministerio de Educaci\'on y 
Ciencia (Spain) under grant FIS2007-66485-C02-02, and the Generalitat de
Catalunya and the Ministerio de Educaci\'on y Ciencia ({\em Programa I3}).


\begin{thebibliography}{10}
\bibitem{correlsign}
We consider the pair correlation of dislocations irrespective of the sign
of their Burgers vector, while in \cite{ZAI-01} dislocations
of the same and different signs were considered.
\end{thebibliography}
\end{document}